\documentclass[aps,pra, twocolumn, noshowpacs, floatfix]{revtex4}

\usepackage{amsfonts}
\usepackage{amssymb}
\usepackage{graphicx}
\usepackage{subfigure}
\usepackage{amsmath}
\usepackage[english]{babel}
\usepackage{color}
\usepackage{epstopdf}
\usepackage{footnote}

\begin{document}

\title{Quantum phase transition of sub-Ohmic spin-boson models: An approach by the multiple Davydov D$_{2}$ \textit{Ansatz}}

\author{Justin Zhengjie Tan$^{1}$, Nengji Zhou$^{2}$ and Yang Zhao$^{1}$\footnote{Electronic address:~\url{YZhao@ntu.edu.sg}}}
\affiliation{$^{1}$\mbox{School of Materials Science and Engineering, Nanyang Technological University, Singapore 639798, Singapore}\\
$^{2}$\mbox{School of Physics, Hangzhou Normal University, Hangzhou 310036, China} \\
}

\begin{abstract}
The ground state properties and quantum phase transitions of sub-Ohmic spin-boson models are investigated using the multiple Davydov D$_{2}$ \textit{Ansatz} in conjunction with the variational principle. Three variants of the model are studied: (i) a single bath with diagonal coupling, (ii) two independent baths with diagonal and off-diagonal couplings, and (iii) a single bath with simultaneous diagonal and off-diagonal couplings. For the purely diagonal model, the multiple Davydov D$_{2}$ \textit{Ansatz} yields critical coupling strengths that are consistent with other methodologies, validating its accuracy and efficiency. In the two-bath model, the competition between diagonal and off-diagonal couplings drives a first-order transition for both symmetric and asymmetric spectral exponents, with von-Neumann entropy showing a continuous peak only under exact symmetry. Finally, for a single bath with simultaneous diagonal and off-diagonal couplings, we demonstrate that a rotational transformation maps the system to an equivalent purely diagonal model, enabling simpler and intuitive physical interpretation and reduced computational complexity.
\end{abstract}
\date{\today}
\maketitle

\section{Introduction}
Simulating quantum systems is essential in many disciplines of physics and chemistry, often complementing experiments to connect theoretical understanding with real-world observations. To accurately model the properties of a quantum system, its mathematical framework must account for environmental interactions, as no physical system can be completely isolated. These environmental effects can lead to phenomena such as altered energy-level interactions in qubits  \cite{TBSBM_v14d}, quantum Zeno effect \cite{quantum_zeno} and realisation of discrete time crystal in an open modulated Dicke model \cite{Gong_DTC,Zhu_DTC}.\\

As a central paradigm of open quantum systems, the spin-boson model (SBM) has made its way into a myriad of fields such as condensed matter physics and physical chemistry with topics including quantum computing \cite{Mag_QC,Vion_QC}, qubit-bath entanglement \cite{Morozov_QBE,McCutcheon_QBE,Wang_QBE} and quantum phase transitions \cite{Guo_QPT,Zhang_QPT,Kirchner_QPT}. Popularized by Anthony J.~ Leggett \cite{Leggett_Chakravarty}, the SBM describes a two level system coupled linearly to a bath of harmonic oscillators which represents the environmental influence. Gapless bosonic baths, characterised by a spectral density function $J(\omega) \propto \omega^s$ adopted after a power-law form, can be classified into three different regimes depending on the value of the spectral exponent $s$: sub-Ohmic ($s<1$) , ohmic ($s=1$) and super-Ohmic ($s>1$) \cite{Wilner, florian, K_Hur, Nacke}. In the super-Ohmic case, the model fails to support any critical fixed points due to the pronounced high-frequency modes. Hence, it always ends up in the delocalized phase and no quantum phase transition exists. Even at the strongest coupling strength, the dynamics turns neither localized nor incoherent, hence the coherent-incoherent dynamical transitions disappear. The system only exhibits damped oscillatory behaviour in such super-Ohmic environment. The special case of $s=1$ corresponds to the ohmic spin-boson model which is also equivalent to the anisotropic Kondo model. The system shows a Kosterlitz-Thouless (KT) type quantum phase transition at the critical coupling $\alpha_c = 1$. The quantum decoherence or maximal entanglement occurs at the Toulouse limit $\alpha=1/2$, whereas it occurs at the transition point in the sub-Ohmic case. A dynamic transition from coherent oscillatory behaviour to incoherent dynamics occurs at the same point of $\alpha=1/2$. \\

\begin{figure}[htp]
	\centering
	\includegraphics[scale=0.25,trim=20 0 0 0]{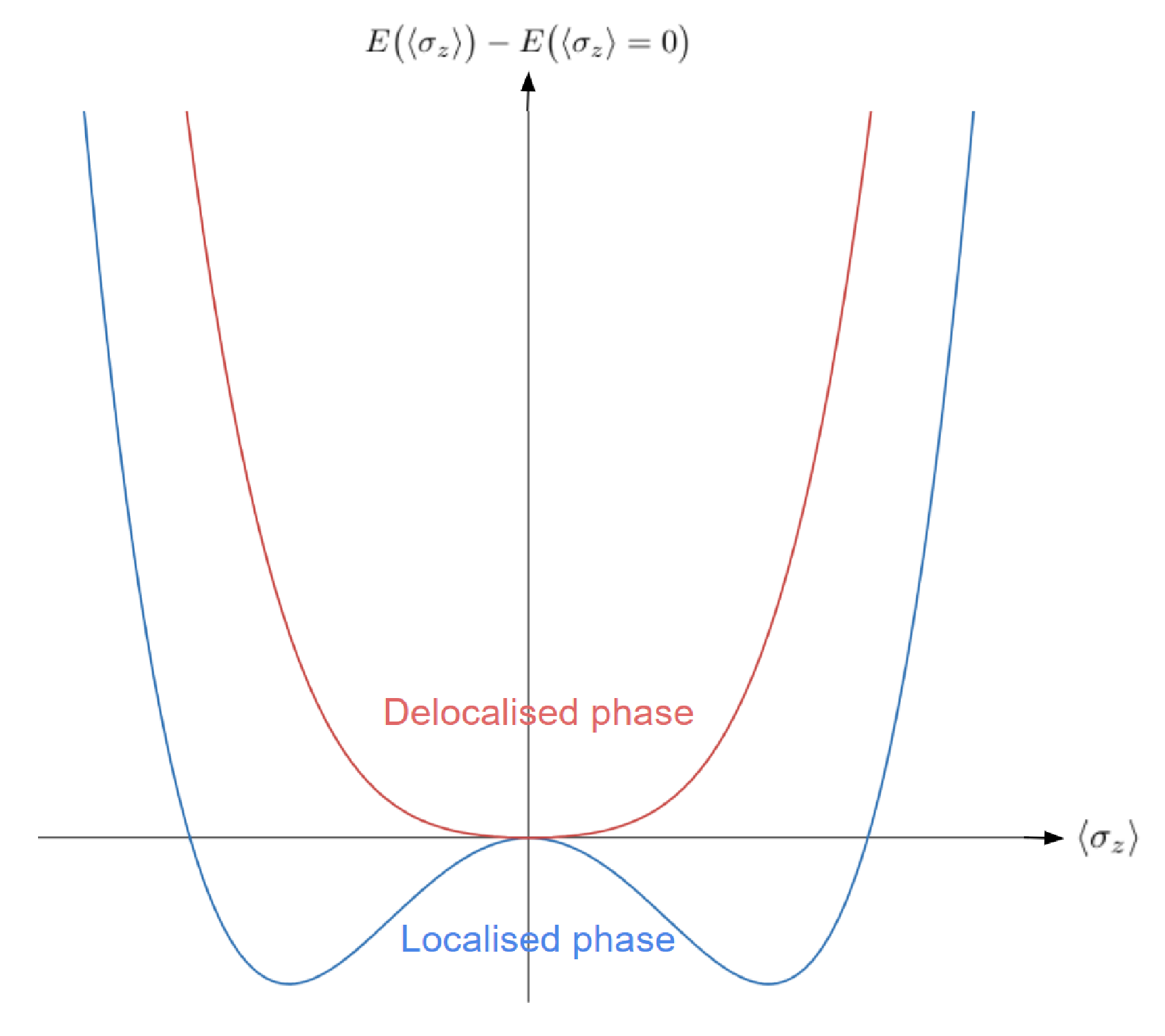} \\
	\caption{Illustration of the ground state energy of the SBM excluding any external bias. A doubly degenerate localised phase is separated from a non-degenerate delocalised phase by a quantum phase transition. The system energy, represented by the y-axis, uses the energy of the system when $\langle \sigma_z \rangle = 0$ as a benchmark.}
	\label{Fig1}
\end{figure}

Most studies on the quantum phase transition in the SBM only consider the diagonal coupling as the inclusion of the off-diagonal coupling makes the identification of the quantum phase transition a challenge. In the sub-Ohmic regime, the SBM with diagonal coupling only is expected to undergo a second order phase transition between a non-degenerate delocalized phase and a doubly degenerate localized phase \cite{Kehrein_Mielke} as depicted in Fig. \ref{Fig1}. With the addition of off-diagonal coupling, the degeneracy of the localized phase vanishes and the transition may no longer be classified as second order \cite{Zhou_Chen_2014,Zhao_Zhao_2014}. In the Ohmic regime, a Kosterlitz-Thouless type phase transition has been well documented and there is no phase transition in the super-Ohmic regime. \\

Despite extensive study, an exact analytical solution still remains elusive which necessitates the use of various numerical methodologies to unveil the ground state properties of the SBM. Variational matrix product states (VMPS) \cite{Guo_QPT}, multiple Davydov D$_1$ \textit{Ansatz} (multi-D$_1$ \textit{Ansatz}) \cite{Zhou_Chen_2014}, numerical renormalization group (NRG) \cite{Bulla_Tong_Vojta_2003,Vojta_Tong_Bulla_2005,Bulla_Lee_Tong_Vojta_2005}, quantum Monte Carlo (QMC) \cite{Winter_Rieger_Vojta_Bulla_2009}, density matrix renormalization group (DMRG) \cite{Wong_Chen_2008}, and exact diagonalization (ED) \cite{Alvermann_Fehske_2009} are among the most popular numerical approaches. In this study, the multiple Davydov D$_2$ \textit{Ansatz} (multi-D$_2$ \textit{Ansatz}) is adopted which has not been used before in this context to the best of the author's knowledge. The multi-D$_2$ \textit{Ansatz} is a sister methodology of the multi-D$_1$ \textit{Ansatz} under the family of the Davydov \textit{Ans\"{a}tze} \cite{Zhao_2023}.

\section{Methodology}
\subsection{Model}
The Hamiltonian of the two-bath spin-boson model can be expressed as

\begin{eqnarray}\label{H_TBSBM}
    \hat{H}_{\text{TBSBM}} &=& \frac{\varepsilon}{2}\sigma_z - \frac{\Delta}{2}\sigma_x + \sum_{k,i}\omega_k b_{k,i}^\dagger b_{k,i} \nonumber\\
    &+&\frac{\sigma_z}{2}\sum_k\lambda_k(b_{k,1}^\dagger + b_{k,1}) \nonumber\\
    &+&\frac{\sigma_x}{2}\sum_k\eta_k(b_{k,2}^\dagger + b_{k,2}),
\end{eqnarray}
where $\varepsilon$ represents the spin bias which describes the influence of an external magnetic field, $\Delta$ signifies the tunnelling constant of the two level system and $b_k (b^{\dagger}_k)$ denotes the boson annihilation (creation) operator of the $k$th mode with frequency $\omega_k$. $\lambda_k$ and $\eta_k$ describes the diagonal coupling amplitude and off-diagonal coupling between the spin and the $k$th boson mode respectively. $\sigma_z$ and $\sigma_x$ are the Pauli matrices. The subscript $i = 1,2$ differentiates the two baths.\\

The two-bath SBM Hamiltonian in Eq. \ref{H_TBSBM} can be reformulated into a single-bath form for convenience as follows

\begin{eqnarray}\label{H_TBSBM_new}
        \hat{H}_{\text{TBSBM}} ^{'} &=& \frac{\varepsilon}{2}\sigma_z - \frac{\Delta}{2}\sigma_x + \sum_{k}\omega_k^{'} b_{k}^{'\dagger} b_{k}^{'} \nonumber\\
    &+&\frac{\sigma_z}{2}\sum_k\lambda_k^{'}(b_{k}^{'\dagger} + b_{k}^{'}) \nonumber\\
    &+&\frac{\sigma_x}{2}\sum_k\eta_k^{'}(b_{k}^{'\dagger} + b_{k}^{'})
\end{eqnarray}
via the transformation

\begin{eqnarray}
    \omega_k^{'} &=&
    \begin{cases}
        \omega_k, & 0 < k \leq N, \\
        \omega_{k-N}, & N < k \leq 2N,
    \end{cases} \nonumber\\
    \lambda_k^{'} &=&
    \begin{cases}
        \lambda_k, & 0 < k \leq N, \\
        0, & N < k \leq 2N,
    \end{cases} \nonumber\\
    \eta_k^{'} &=&
    \begin{cases}
        0, & 0 < k \leq N, \\
        \eta_{k-N}, & N < k \leq 2N,
    \end{cases} \nonumber\\
    b_k^{'} &=&
    \begin{cases}
        b_{k,1}, & 0 < k \leq N, \\
        b_{k-N,2}, & N < k \leq 2N,
    \end{cases}
\end{eqnarray}
where $N$ refers to the number of effective bath modes for both the diagonal and off-diagonal baths. Planck's constant $\hbar$ is set to unity to simply notations such that model parameters ($\varepsilon, \Delta, \lambda_k, \eta_k$ and $\omega_k$) are dimensionless. \\

Applying the coarse-grained treatment based on the Wilson energy mesh, the values of $\omega_k$, $\lambda_k$ and $\eta_k$ can be extracted from the diagonal and off-diagonal continuous spectral density functions, represented by $J_z(\omega)$ and $J_x(\omega)$ respectively,

\begin{eqnarray}
    J_z(\omega) &=& \sum_k \lambda_k^2 \delta (\omega-\omega_k) = 2\alpha \omega_c^{1-s}\omega^s \\
    J_x(\omega) &=& \sum_k \eta_k^2 \delta (\omega-\omega_k) = 2\beta \omega_c^{1-\bar{s}}\omega^{\bar{s}},
\end{eqnarray}
where $\alpha$ and $\beta$ denote the dimensionless coupling strengths, and the two boson baths are characterized by the spectral exponents $s$ and $\bar{s}$, accounting for the diagonal and off-diagonal coupling respectively. The high frequency cutoff $\omega_c$ is set to unity throughout this paper.  A logarithmic discretization procedure is adopted by dividing the phonon frequency domain $[0,\omega_c]$ into $N$ intervals $[\Lambda^{-(k+1)},\Lambda^{-k}]\omega_c$ where $k=0,1,...,N-1$. Thereby, the parameters $\omega_k$, $\omega_{k-N}$, $\lambda_k$ and $\eta_{k-N}$ from Eq. \ref{H_TBSBM_new} can be obtained as follows,

\begin{eqnarray}
    \lambda_k^2 &=& \int_{\Lambda^{-k-1}\omega_c}^{\Lambda^{-k}\omega_c} dx J_z(x) \hspace{1cm}~=~~ \frac{2\alpha \omega_c^{1-s}}{s+1}\omega^{s+1},\\
    \omega_k &=& \lambda_k^{-2} \int_{\Lambda^{-k-1}\omega_c}^{\Lambda^{-k}\omega_c} dx J_z(x)x ~~=~ \frac{2\alpha\omega_c^{1-s}}{\lambda_k^2 (s+2)}\omega^{s+2} ,\\
    \eta_{k-N}^2 &=& \int_{\Lambda^{-k-1}\omega_c}^{\Lambda^{-k}\omega_c} dx J_x(x) \hspace{1cm}~=~~ \frac{2\beta \omega_c^{1-\bar{s}}}{\bar{s}+1}\omega^{\bar{s}+1} ,\\
    \omega_{k-N} &=& \eta_{k-N}^{-2} \int_{\Lambda^{-k-1}\omega_c}^{\Lambda^{-k}\omega_c} dx J_x(x)x ~=~ \frac{2\beta\omega_c^{1-\bar{s}}}{\eta_{k-N}^2 (\bar{s}+2)}\omega^{\bar{s}+2}. \nonumber\\
\end{eqnarray}

It should be noted that infinite bath modes are considered via the integration of the continuous spectral density $J(\omega)$, although the number of effective bath modes $N$ is finite. In the continuum limit, the logarithmic discretization factor $\Lambda \to 1$. Therefore, $\Lambda = 1.05$ is fixed throughout this study, unless specified otherwise, to create a highly dense spectrum for an accurate description of the SBM ground state. Additionally, $\Lambda = 1.05$ has been found to be sufficient to attain reliable converged results in previous studies. \\

\subsection{Multiple Davydov D$_2$ \textit{Ansatz}}
The multi-D$_2$ \textit{Ansatz} with multiplicity $M$ can be thought of as linear superpositions of $M$ copies of single Davydov D$_2$ Ansatz \cite{Zhao_Sun_Chen_Gelin_2021}. Theoretically, numerically ``exact" solutions can be obtained with a sufficiently high multiplicity. This Ansatz has been used to study both dynamic and static properties of various many-body systems, producing results that are both numerical efficient and accurate in a wide parameter range in the presence of diagonal and off-diagonal system-bath coupling \cite{Zhou_Huang_zhu,Huang_Chen_Zhou,LZT_2018,Jaku,24d_CIM}. In this work, the multi-D$_2$ \textit{Ansatz} is used to precisely describe the simultaneous diagonal and off-diagonal spin-bath coupling. The variational trial state of Eq. (2) can be introduced as a systematic coherent-state expansion of the ground-state wave function, expressed as
\begin{eqnarray}\label{eq:MD2}
	|{\Psi^M_{D_2}}\rangle &=& |+\rangle \sum_{m=1}^M A_m \exp \Big[ \sum_{k=1}^{N_{total}} f_{m,k} (b_k^{'\dagger}-b_k^{'})\Big] |0\rangle_{ph} \nonumber\\
	&+& |-\rangle \sum_{m=1}^M B_m \exp \Big[ \sum_{k=1}^{N_{total}} f_{m,k} (b_k^{'\dagger}-b_k^{'})\Big] |0\rangle_{ph} ~, \nonumber\\
\end{eqnarray}

where $M$ represents the number of coherent superposition states and $N_{total}$ represents the total number of effective bath modes, which is $N$ in the SBM and $2N$ in the two-bath SBM. Here, $|+\rangle (|-\rangle)$ is the spin-up (spin-down) state and $|0\rangle_{ph}$ is the phonon bath vacuum state. $A_m$ and $B_m$ are the occupation amplitudes of $|+\rangle$ and $|-\rangle$, respectively. For multi-D$_2$ \textit{Ansatz}, both spin states share the same  bosonic displacements $f_{m,k}$. Lastly, $m$ and $k$ denote the $m$th Gaussian superposition and the $k$th boson mode, respectively.\\

The ground state of the system $|{\Psi^M_{D_2}}\rangle$ can be calculated by minimising the energy in the form of $E = \mathcal{H}/\mathcal{N}$, after deriving the Hamiltonian expectation $\mathcal{H} = \langle{\Psi^M_{D_2}}|\hat{H}|{\Psi^M_{D_2}}\rangle$ and the norm of the wave function $\mathcal{N} = \langle{\Psi^M_{D_2}}|{\Psi^M_{D_2}}\rangle$.

\begin{eqnarray}
    \mathcal{H} &=& \sum_{m,n}(A_mB_n+B_mA_n)F_{m,n}\Bigg[-\frac{\Delta}{2}+\sum_k \frac{\eta_k}{2}(f_{m,k}+f_{n,k}) \Bigg] \nonumber\\
    &+& \sum_{m,n}A_mA_nF_{m,n} \nonumber\\
    &\times &\Bigg\{\frac{\varepsilon}{2} + \sum_k \Big[\omega_kf_{m,k}f_{n,k} + \frac{\lambda_k}{2}(f_{m,k}+f_{n,k}) \Big] \Bigg\} \nonumber\\
    &+& \sum_{m,n}B_mB_nF_{m,n} \nonumber\\
    &\times &\Bigg\{-\frac{\varepsilon}{2} + \sum_k \Big[\omega_kf_{m,k}f_{n,k} - \frac{\lambda_k}{2}(f_{m,k}+f_{n,k}) \Big] \Bigg\}
\end{eqnarray}

and

\begin{eqnarray}
   \mathcal{N} = \sum_{m,n}(A_mA_nF_{m,n} + B_mB_nF_{m,n})
\end{eqnarray}

where $F_{m,n}$ refers to the Debye-Waller factor which is expressed as

\begin{eqnarray}
    F_{m,n}=\exp\bigg[-\frac{1}{2}\sum_k (f_{m,k}-f_{n,k})^2\bigg].
\end{eqnarray}

\begin{table*}[tbp]
\caption{Using various methodologies, the critical coupling $\alpha_c$ is shown for a wide range of the spectral exponent $s$ in the sub-Ohmic SBM. The table is imported from Ref. \cite{Shen_Zhou_2023} and the results of NRG, QMC and DMRG is obtained via plot digitising techniques with the respective error bar listed as the maximum inaccuracy of the readout in brackets. The novel multi-D$_2$ \textit{Ansatz} results are posted in the final column of the table. The following parameters are fixed for the multi-D$_2$ \textit{Ansatz} column: $\Delta=0.1$, $\Lambda=1.05$, $M=8$, $N=400$.}
\begin{center}

\begin{tabular}{p{2cm}p{2cm}p{2cm}p{2cm}p{2cm}p{2cm}p{2cm}p{2cm}}
\hline
\hline
$s$ & NRG & QMC & VMPS & DMRG & Single ~~~~~~~~~polaron & multi-D$_1$ \textit{Ansatz} & multi-D$_2$ \textit{Ansatz}\\
\hline
0.1 & 0.008(1) & 0.0076(3) & -         & 0.0074(2) & 0.0065 & 0.00795(1) & 0.0084\\
0.2 & 0.018(1) & 0.0175(2) & 0.0175367 & 0.0162(2) & 0.0168 & 0.01806(3) & 0.0182\\
0.3 & 0.035(2) & 0.0350(5) & 0.0346142 & 0.0332(5) & 0.0316 & 0.03470(6) & 0.0349\\
0.4 & 0.064(2) & 0.0604(8) & 0.0605550 & 0.058(1)  & 0.0519 & 0.0604(2)  & 0.0605\\
0.5 & 0.106(2) & 0.098(1)  & 0.0990626 & 0.099(1)  & 0.0784 & 0.0977(3)  & 0.0981\\
\hline
\hline
\end{tabular}
\end{center}
\label{table1}
\end{table*}

A set of self-consistency equations are then deduced by the Lagrange multiplier method while considering the constraint $\mathcal{N} \equiv 1$,

\begin{eqnarray}
    \frac{\partial\mathcal{H}}{\partial x_i} - E\frac{\partial\mathcal{N}}{\partial x_i} = 0,
\end{eqnarray}

with respect to variational parameters $x_i$, where $i = 1,2,3,...,M(2N_{total}+2)$. Ultimately, it follows that the iterative equations of the variational parameters can be shown as

\begin{eqnarray}
    A_n^* &=& \frac{2\sum_m B_mF_{n,m}d_{n,m}+2\sum^{m\neq n}_m A_mF_{m,n}(a_{m,n}-E)}{2(E-a_{n,n})}, \nonumber\\ \\
    B_n^* &=& \frac{2\sum_m A_mF_{m,n}d_{m,n}+2\sum^{m\neq n}_m B_mF_{m,n}(b_{m,n}-E)}{2(E-b_{n,n})}, \nonumber\\
\end{eqnarray}
\begin{eqnarray}
    &f_{m,k}^{*}& = \frac{\sum_{n}(A_nB_m+B_nA_m)F_{n,m}\Big[2d_{n,m}f_{n,k}+\eta_k \Big]}{2(A_m^2+B_m^2)(E-\omega_k)-2(A_m^2a_{m,m}+B_m^2b_{m,m})} \nonumber \\
    &+& \frac{\sum_{n}^{n \neq m}A_nA_mF_{n,m}\Big[2a_{n,m}f_{n,k} + 2\omega_kf_{n,k}+\lambda_k -2f_{n,k}E\Big]}{2(A_m^2+B_m^2)(E-\omega_k)-2(A_m^2a_{m,m}+B_m^2b_{m,m})} \nonumber \\
    &+& \frac{\sum_{n}^{n \neq m}B_nB_mF_{n,m}\Big[2b_{n,m}f_{n,k} + 2\omega_kf_{n,k}-\lambda_k -2f_{n,k}E \Big]}{2(A_m^2+B_m^2)(E-\omega_k)-2(A_m^2a_{m,m}+B_m^2b_{m,m})} \nonumber\\
    &+& \frac{(A_m^2-B_m^2)\lambda_k}{2(A_m^2+B_m^2)(E-\omega_k)-2(A_m^2a_{m,m}+B_m^2b_{m,m})}, \nonumber\\
\end{eqnarray}

where $a_{m,n}$, $b_{m,n}$ and $d_{m,n}$ denote

\begin{eqnarray}\label{Eq. x_i}
    a_{m,n} &=& \frac{\varepsilon}{2}+\sum_k \Big[\omega_kf_{m,k}f_{n,k}+\frac{\lambda}{2}(f_{m,k}+f_{n,k}) \Big], \\
    b_{m,n} &=& -\frac{\varepsilon}{2}+\sum_k \Big[\omega_kf_{m,k}f_{n,k}-\frac{\lambda}{2}(f_{m,k}+f_{n,k}) \Big], \nonumber \\ \\
    d_{m,n} &=& -\frac{\Delta}{2}+\sum_k\frac{\eta_k}{2}(f_{m,k}+f_{n,k}).
\end{eqnarray}

The relaxation iteration procedure is adopted, updating the variational parameter by $x_i^{'} = x_i + t*(x_i^{*}-x_i)$ where $x_i^{*}$ represents $A_n^*$, $B_n^*$ and $f^*_{m,k}$ while $t = 0.1$ is the relaxation factor. The termination criterion of the iteration algorithm is imposed as max$\{x_i^{*}-x_i\}<10^{-11}$ over all the variational parameters. More than ten random
initial states are calculated to reduce statistical noise for each set of model parameters ($\alpha, \beta, \Delta, \Lambda, \varepsilon$). Additionally, simulated annealing algorithm is also employed to escape from metastable states.\\

Thereafter, the spin magnetisation $\langle \sigma_z \rangle$, the spin coherence $\langle \sigma_x \rangle$ and the von-Neumann entropy (also known as entanglement entropy) $S_{v-N}$ that is used to characterise the entanglement between the spin and its environmental bath can be introduced as

\begin{eqnarray}\label{Eq. gs properties}
    \langle \sigma_z \rangle &=& \frac{\sum_{m,n} \big[ F_{m,n}(A_mA_n - B_mB_n) \big]}{\sum_{m,n}\big[F_{m,n}(A_mA_n + B_mB_n)\big]}, \\
     \langle \sigma_x \rangle &=& \frac{\sum_{m,n} \big[ F_{m,n} (A_mB_n + B_mA_n) \big]}{\sum_{m,n}\big[F_{m,n}(A_mA_n + B_mB_n)\big]}, \\
     S_{v-N} &=& -\omega_+ \log \omega_+ ~-~ \omega_- \log \omega_-,
\end{eqnarray}
where $\omega_{\pm} = \big[1 \pm \sqrt{\langle \sigma_z \rangle^2 + \langle \sigma_y \rangle^2 + \langle \sigma_x \rangle^2} ~\big]/2$.

\section{Results and Discussion}

\subsection{Diagonal coupling only ($\beta = 0$)}
This section aims to evaluate the numerical accuracy of the multi-D2 ansatz in comparison with other numerical methods, as summarized in Table \ref{table1}. \\

The transition points $\alpha_c$ of various spectral exponent $s\in[0.1,0.5]$ is almost indistinguishable between multi-D$_1$ \textit{Ansatz} and multi-D$_2$ \textit{Ansatz}. The difference between the two sister trial states is that for multi-D$_1$ \textit{Ansatz}, the boson displacements for the spin up and spin down states are unique whereas the boson displacements are identical for both states in multi-D$_2$ \textit{Ansatz}. The multi-D$_1$ trial state has the form:

\begin{eqnarray}\label{eq:MD1}
	|{\Psi^M_{D_1}}\rangle &=& |+\rangle \sum_{m=1}^M A_m \exp \Big[ \sum_{k=1}^N f_{m,k} (b_k^\dagger-b_k)\Big] |0\rangle_{ph} \nonumber\\
	&+& |-\rangle \sum_{m=1}^M B_m \exp \Big[ \sum_{k=1}^N g_{m,k} (b_k^\dagger-b_k)\Big] |0\rangle_{ph} \nonumber\\
\end{eqnarray}
Upon closer inspection, it can be shown that the multi-D$_1$ \textit{Ansatz} can be derived from the multi-D$_2$ \textit{Ansatz} under special circumstances: 1) the multi-D$_2$ \textit{Ansatz} with even multiplicity, 2) $A_m = 0$ for even $m$  and $B_m = 0$ for odd $m$, can be simply reduced to the multi-D$_1$ \textit{Ansatz} with half the multiplicity \cite{Zhao_2023}. In Table 1, for the multi-D$_1$ \textit{Ansatz} (the multi-D$_2$ \textit{Ansatz}), the multiplicity is 8 (12), the number of modes is 430 (400) and the number of variational parameters is 6896 (4824). Despite having lesser variational parameters, largely because the multiplicity of multi-D$_2$ \textit{Ansatz} is less than twice than the multiplicity in multi-D$_1$ \textit{Ansatz}, the results yielded by the two trial states converge, which suggests higher numerical efficiency and shorter CPU time for the multi-D$_2$ \textit{Ansatz}. Moreover, the variational approach based on the multi-D$_2$ \textit{Ansatz}  have also been shown to be valid in describing the ground-state transition in the Ohmic spin-boson model, a numerically challenging problem since the transition has been predicted of the Kosterlitz-Thouless type, thereby lending support to the versatility of the multi-D$_2$ \textit{Ansatz} \cite{bera2014}. Furthermore, the results from the multi-D$_2$ \textit{Ansatz} is in good agreement with other methodologies, which lends support to the notion that the multi-D$_2$ \textit{Ansatz} should be a valid and strong competitor among various state-of-the-art numerical methods. \\

On a tangent, in the research of the Holstein polaron, the multi-D$_1$ \textit{Ansatz} is typically recruited for studying Hamiltonians containing only diagonal coupling as the phonon displacements are dependent on both the sites and momentum which can yield numerically accurate results. However, in the presence of off-diagonal coupling, the multi-D$_2$ \textit{Ansatz} is usually used as the phonon displacement is site-independent which is more suitable in delocalised situations in the presence of nonlocal or off-diagonal coupling \cite{Zhou_Huang_zhu, ZhouChenHuang}. However, if the multiplicity is sufficiently high, the distinction between the two trial states vanishes as they both approach the numerically ``exact" solution. Following this tradition, multi-D$_2$ \textit{Ansatz} is chosen for the next section as the Hamiltonian contains off-diagonal coupling even though multi-D$_1$ \textit{Ansatz} is a viable option as well. \\

In molecular excitonic systems, the variational polaron transformation is employed to study the energy transfer and dynamics of the electronic coupling and the vibronic interactions \cite{Reppert,Zimanyi}. Both of which are competitive effects with the former supporting delocalization and the latter driving localization, which is very similar to the role that the off-diagonal coupling and diagonal coupling plays in the SBM respectively. In vibrational-excitonic systems at zero temperature, the minimization of the free energy is identical to the minimization of the ground state energy in this work. The single polaron \textit{Ansatz}, which is a generalisation of the Silbey-Harris polaron \textit{Ansatz}, and arising from the variational polaron transformation has an analytical solution \cite{single_polaron} in the deep sub-Ohmic SBM in the form of 

\begin{eqnarray}
\alpha_c = \frac{\sin(\pi s)e^{-s/2}}{2\pi (1-s)}\Big( \frac{\Delta}{\omega_c} \Big)^{1-s}.
\end{eqnarray}

The critical transition points of the analytical solution from the single polaron \textit{Ansatz}, included in Table \ref{table1}, is not in close agreement with other numerical methods. The trial wave function of the single polaron \textit{Ansatz} is not complex enough to capture the sophisticated effects of quantum entanglement and quantum fluctuations in the bosonic bath which is imperative to the quantum phase transition. Numerical methods like the multi-D$_2$ \textit{Ansatz} which offers more variational parameters in the trial wave function has greater flexibility to mimic the ``exact" ground state wave function.   

\subsection{Diagonal and off-diagonal coupling} \label{ss_B}

In this section, $\Delta$ and $\varepsilon$ are set to $0$, with the dimensionless diagonal coupling $\alpha$ fixed at $0.02$. The multiplicity $M=12$, number of modes $N=20$ and the discretization factor $\Lambda=2$.\\

\begin{figure}[htp]
	\centering
	\includegraphics[scale=0.33]{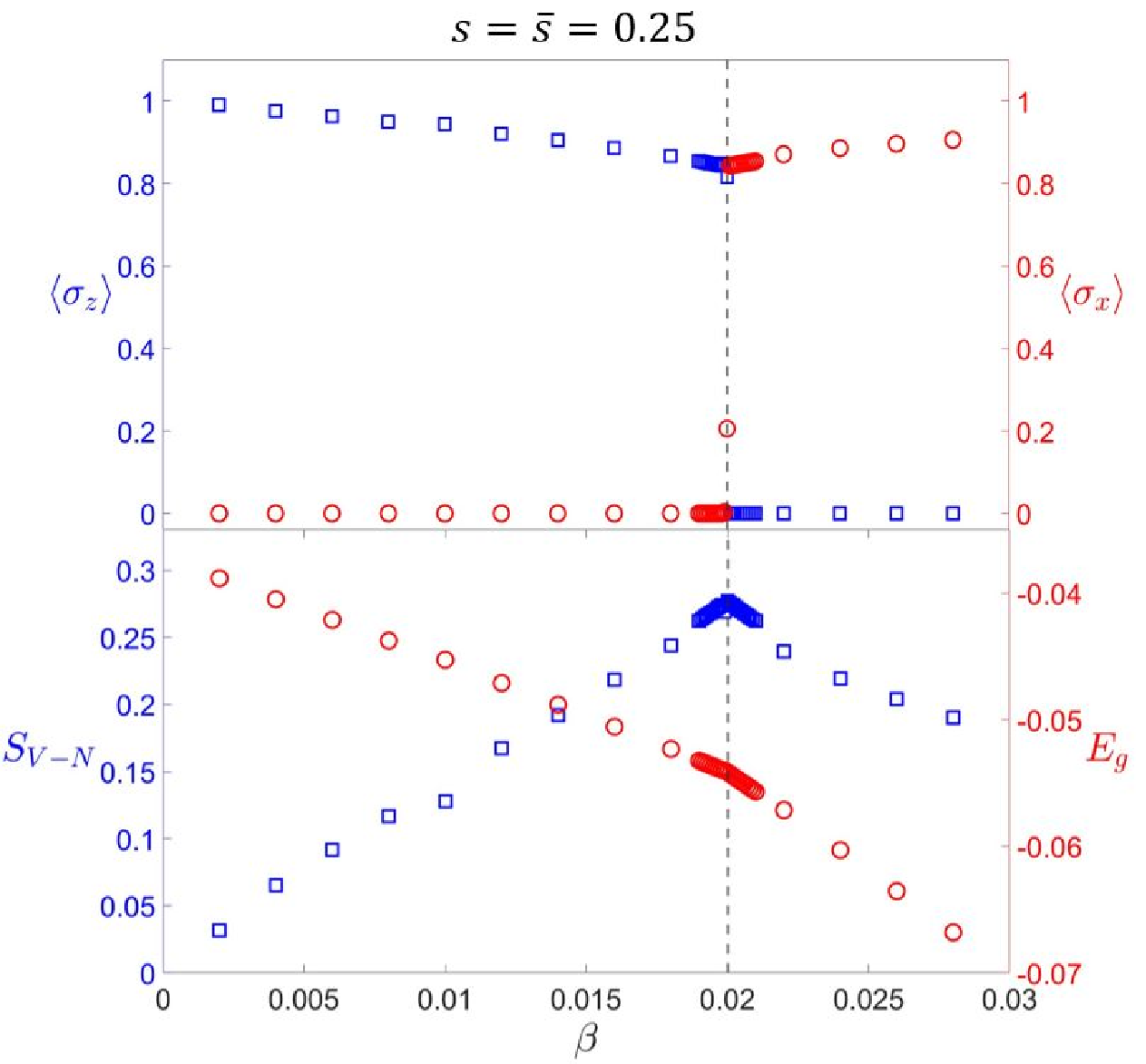} \\
	\caption{Top plot: $\langle\sigma_z\rangle$ (blue squares) and $\langle\sigma_x\rangle$ (red circles) as a function of $\beta$. Bottom plot: von-Neumann entropy $S_{v-N}$ (blue squares) and ground state energy $E_g$ (red circles) as a function of $\beta$. This is obtained with $s=\bar{s}=0.25$, $\alpha=0.02$, $\varepsilon=\Delta=0$. The dotted line denotes the transition point $\beta_c=0.02$.}
	\label{Fig2}
\end{figure}

\begin{figure}[htp]
	\centering
	\includegraphics[width=0.9\linewidth, height=6cm]{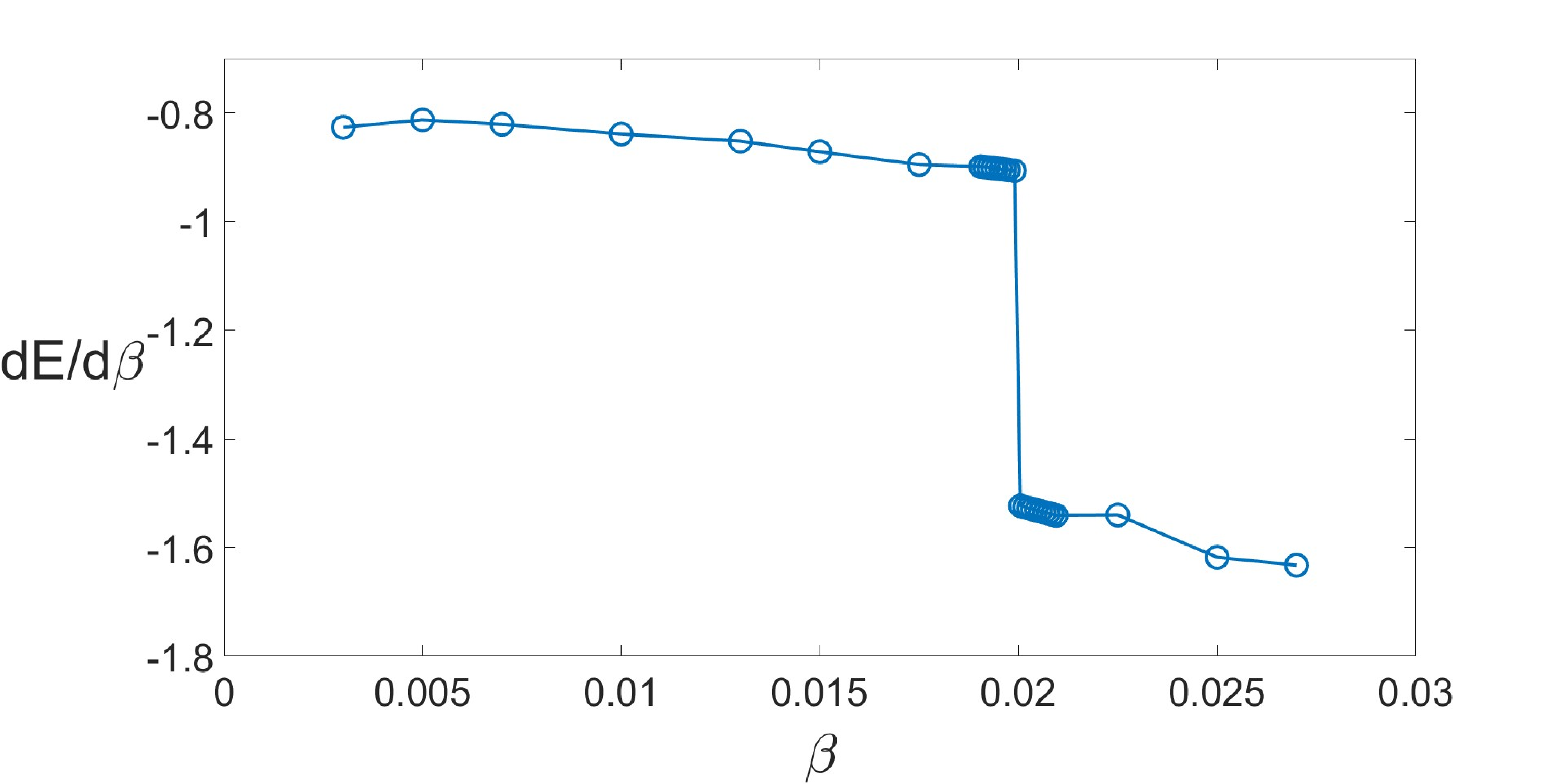} \\
	\caption{The derivative of the ground-state energy $E_g$ is displayed with respect to the off-diagonal coupling strength $\beta$. This is obtained with $s=\bar{s}=0.25$, $\alpha=0.02$, $\varepsilon=\Delta=0$.}
	\label{energygrad}
\end{figure}

 First, the case where $s=\bar{s}=0.25$ is investigated. In Fig. \ref{Fig2}, the top panel reveals $\langle \sigma_z \rangle$ and $\langle \sigma_x \rangle$ with $\beta$ ranging from $0$ to $0.028$. For simplicity, only one branch of the doubly degenerate ground state is shown as $\langle \sigma_z \rangle$ and $\langle \sigma_x \rangle$ tends to bifurcate before and after the transition point respectively. The other degenerate branch can be obtained by projecting the operator $P_z$ or $P_x$ onto the ground state, where $P_z=\sigma_z e^{i\pi \Sigma_n b_{n,2}^{\dagger} b_{n,2}}$ and $P_x=\sigma_x e^{i\pi \Sigma_n b_{n,1}^{\dagger} b_{n,1}}$. The bottom panel presents the von-Neumann entropy $S_{v-N}$ on the left axis and the ground state energy on the right axis. The abrupt discontinuous changes in $\langle \sigma_z \rangle$ and $\langle \sigma_x \rangle$ at the transition point $\beta_c=0.02$ are features of a first order phase transition. The diagonal coupling and off-diagonal coupling are competitive effects. The diagonal coupling causes the spin to be localised, adopting a finite $\langle\sigma_z\rangle$ value whereas the off-diagonal coupling induces the spin to flip between the two states, entering a state of superposition where $\langle\sigma_z\rangle=0$. With such highly symmetric conditions, when $\alpha>\beta$, the diagonal coupling is more dominant than the off-diagonal coupling and $\langle\sigma_z\rangle$ takes on a finite value. Conversely, when $\beta>\alpha$, the off-diagonal coupling is stronger, causing $\langle\sigma_x\rangle$ to be finite and $\langle\sigma_z\rangle$ to be in a superposition. When $\alpha=\beta$ at the transition point, the diagonal and off-diagonal couplings are exactly balanced causing the spin to be maximally entangled by the two baths as the spin is equally influenced by both orthogonal baths. This is counter-intuitive as the entropy is continuous despite being a first order transition which only occurs under such symmetric conditions. The derivative of the ground state energy $\partial E_g/ \partial \beta$, illustrated in Fig. \ref{energygrad}, shows a distinct discontinuity at the critical transition point. This observation provides additional evidence that the phase transition in the two-bath SBM with symmetric ohmcity is of first order nature. \\
 
\begin{figure}[htp]
	\centering
	\includegraphics[scale=0.33]{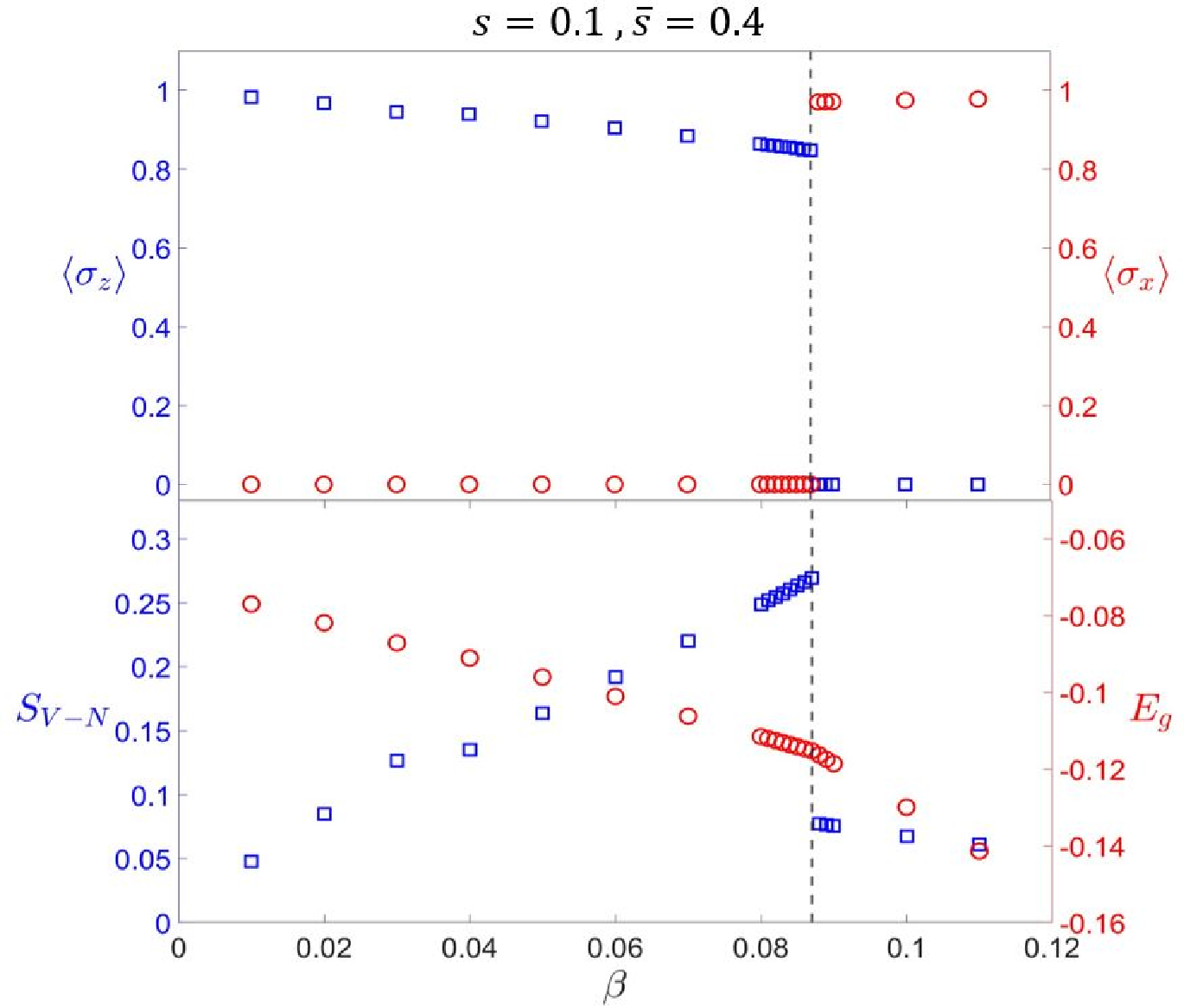} \\
	\caption{Top plot: $\langle\sigma_z\rangle$ (blue squares) and $\langle\sigma_x\rangle$ (red circles) as a function of $\beta$. Bottom plot: $S_{v-N}$ (blue squares) and ground state energy $E_g$ (red circles) as a function of $\beta$. This is obtained with $s=0.1$, $\bar{s}=0.4$, $\alpha=0.02$, $\varepsilon=\Delta=0$. The dotted line denotes the transition point $\beta_c=0.087$.}
	\label{Fig3}
\end{figure}

In the case of $s\neq\bar{s}$, the results are illustrated in Fig. \ref{Fig3}, where $s=0.1$ and $\bar{s}=0.4$. At the transition point $\beta_c=0.087$, there is a discontinuous change of $S_{v-N}$ accompanied by steep changes in $\langle\sigma_z\rangle$ and $\langle\sigma_x\rangle$ values. These are trademark features of a first order transition. With different spectral exponents, the dimensionless coupling strength of $\alpha$ and $\beta$ can be renormalised by a factor of $\frac{\omega^{s+1}}{s+1}$ and $\frac{\omega^{\bar{s}+1}}{\bar{s}+1}$ respectively. Selecting $s=0.1$ and $\bar{s}=0.4$, leads to an increase in the effective dimensionless diagonal coupling strength and a decrease in effective dimensionless off-diagonal coupling strength. This causes the critical point to shift towards a higher $\beta$ value. Other than the nature of the phase transition, the physical picture is largely similar to the case of $s=\bar{s}$.

\subsection{Rotational Theory}

For the rotation theory to function, the SBM should only have one boson bath with each individual mode coupled diagonally and off-diagonally simultaneously to the spin. The Hamiltonian can be expressed as

\begin{eqnarray}
    \hat{H}&=&\frac{\varepsilon}{2}\sigma_z - \frac{\Delta}{2}\sigma_x + \sum_k\omega_kb_k^\dagger b_k \nonumber\\
    &+&\frac{\sigma_z}{2}\sum_k \lambda_k(b_k^\dagger + b_k) \nonumber\\
    &+&\frac{\sigma_x}{2}\sum_k \eta_k(b_k^\dagger + b_k)
\end{eqnarray}

Employing a rotation operator, the SBM from above can be rotated as shown below

\begin{eqnarray}
    \hat{H}_{rot} &=& \langle e^{i\frac{\theta}{2}\sigma_y} | \hat{H} | e^{-i\frac{\theta}{2}\sigma_y} \rangle \nonumber\\
    &=& \frac{\tilde{\varepsilon}}{2}\sigma_z - \frac{\tilde{\Delta}}{2}\sigma_x + \sum_k\tilde{\omega}_kb_k^\dagger b_k \nonumber\\
    &+&\frac{\sigma_z}{2}\sum_k\tilde{\lambda}_k(b_k^\dagger + b_k) \nonumber\\
    &+&\frac{\sigma_x}{2}\sum_k \tilde{\eta}_k(b_k^\dagger + b_k)
\end{eqnarray}

where

\begin{eqnarray}
    \tilde{\varepsilon} &=& \varepsilon \cos\theta - \Delta\sin\theta \\
    \tilde{\Delta} &=& \varepsilon\sin\theta + \Delta\cos\theta \\
    \tilde{\lambda} &=& \lambda\cos\theta + \eta\sin\theta \\
    \tilde{\eta} &=& -\lambda\sin\theta + \eta\cos\theta \\
    \tilde{\omega} &=& \omega \Big(\frac{\tilde{\alpha}}{\tilde{\lambda}} \Big) / \Big(\frac{\alpha}{\lambda} + \frac{\beta}{\eta} \Big)
\end{eqnarray}

With this transformation, the rotated Hamiltonian
$\hat{H}_{rot}$ has a new modified spin bias $\tilde{\varepsilon}$, tunnelling constant
$\tilde{\Delta}$, diagonally coupling $\tilde{\lambda}$ and off-diagonally coupling $\tilde{\eta}$ term which can all be expressed in terms of its pre-rotation terms. From here, two conditions are imposed,
\begin{enumerate}
    \item  Consider that there is no spin bias post-rotation ($\tilde{\varepsilon} = 0$), which leads to the relationship $\tan\theta=\varepsilon/\Delta$.
    \item Assume that there is no off-diagonal coupling post-rotation ($\tilde{\eta}=0$), the relationship of $\tan\theta = \eta/\lambda$ can be obtained.
    \begin{eqnarray}
        \frac{\varepsilon}{\Delta}=\tan\theta = \frac{\eta}{\lambda} = \sqrt{\beta/\alpha}
    \end{eqnarray}
\end{enumerate}

The rest of the parameters are set as follows: $s=\bar{s}=0.3$, $\theta=\pi/3$, $3\alpha=\beta$, $\Delta=0.1$, $\varepsilon=0.1\sqrt{3}$, $\tilde{\Delta}=0.2$, $M=14$, $N=50$ and $\Lambda=1.5$. With these parameters chosen, it is evident that $4\alpha = \tilde{\alpha}$. \\

\begin{figure}[htp]
	\centering
	\includegraphics[width=0.9\linewidth]{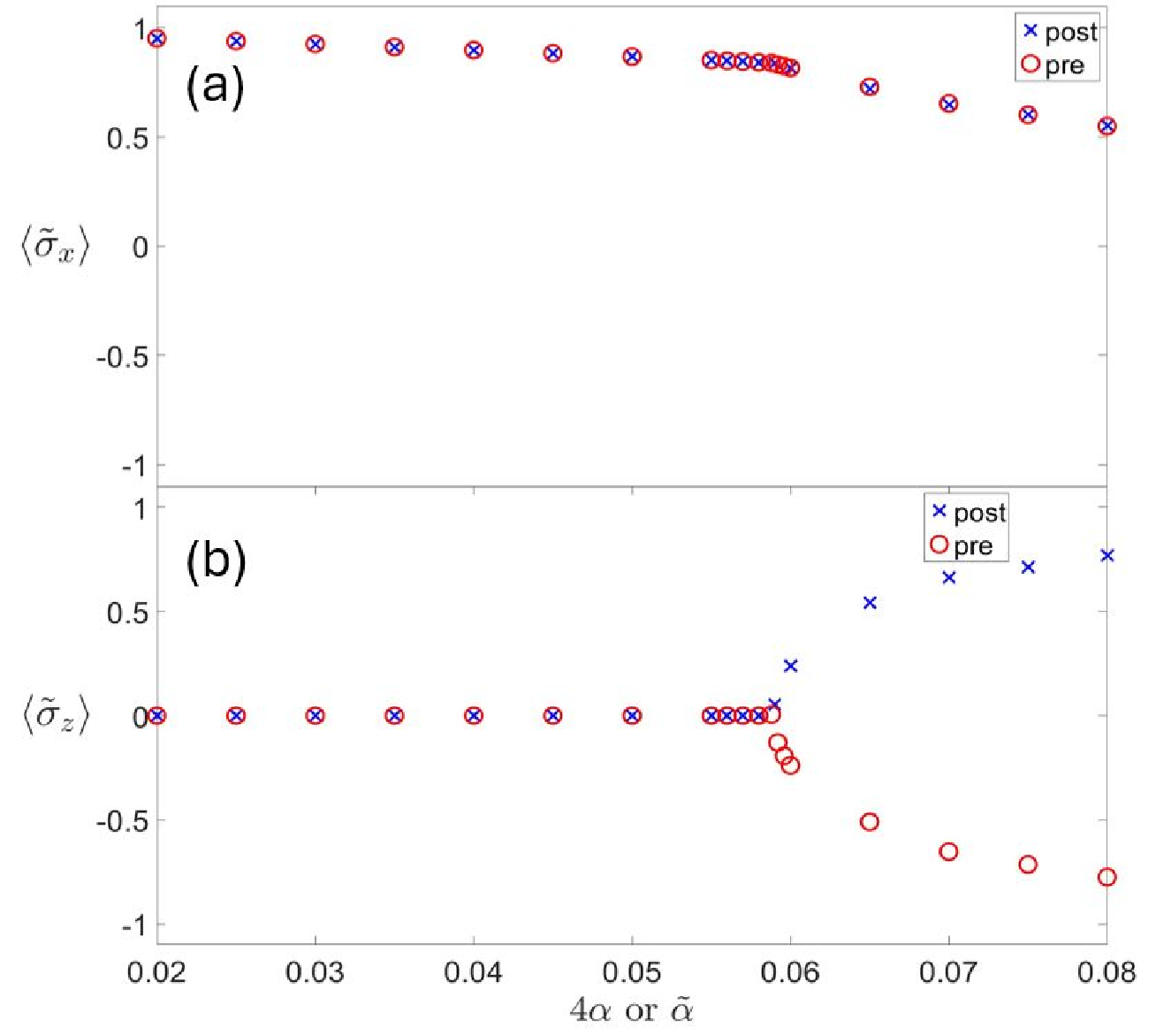} \\
	\caption{(a) $\langle \tilde{\sigma}_z \rangle$ and (b) $\langle \tilde{\sigma}_x \rangle$ as a function of $\tilde{\alpha}$ or $4\alpha$. The red legend represents the expectation values obtained from the pre-rotated Hamiltonian in Eq.~(19) and rotated into the post-rotation frame. The blue legend represents the expectation values obtained from the post-rotated Hamiltonian in Eq. 20.}
	\label{Fig4}
\end{figure}

\begin{figure*}[htp]
	\centering
	\includegraphics[width=0.7\textwidth, height=0.3\textheight]{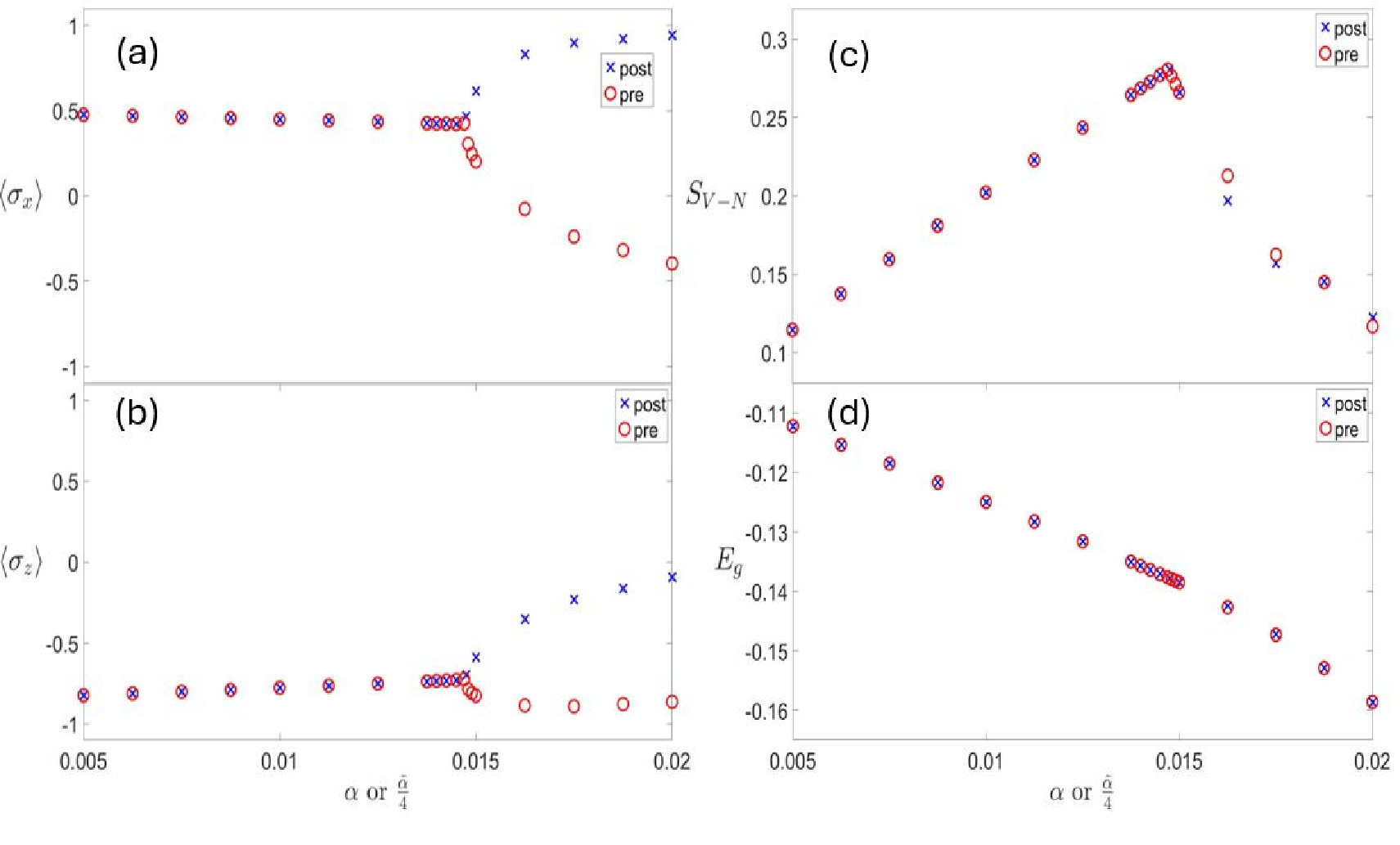} \\
	\caption{(a) $\langle \tilde{\sigma}_z \rangle$ and (b) $\langle \tilde{\sigma}_x \rangle$ (bottom panel) as a function of $\tilde{\alpha}/4$ or $\alpha$. The red legend represents the expectation values obtained from the pre-rotated Hamiltonian in Eq.~(19). The blue legend represents the expectation values obtained from the post-rotated Hamiltonian in Eq.~(20). and rotated into the pre-rotation frame. (c) The von-Neumann entropy and (d) the ground state energy obtained from Eq.~(19) (red legend) and Eq.~(20) (blue legend).}
	\label{Fig5}
\end{figure*}

The expectation values of $\sigma_z$ and $\sigma_x$ in the pre-rotation frame can be converted into the post-rotation frame by substituting $\theta=-\pi/3$ into the following rotation equations $\tilde{\sigma}_z = \sigma_z\cos\theta - \sigma_x\sin\theta$ and $\tilde{\sigma}_x = \sigma_z\sin\theta + \sigma_x\cos\theta$. In the post-rotation frame, there is zero bias and only diagonal coupling. In such cases, it is known that as $\alpha \to 0$, $\langle \sigma_z \rangle \to 0$ and $\langle \sigma_x \rangle \to 1$. At $\tilde{\alpha}_c=0.059$, a second-order transition occurs and $\langle \sigma_z \rangle$ starts to bifurcate in a symmetric fashion revealing the degenerate ground states. Following which, as $\alpha \to \infty$, $|\langle \sigma_z \rangle| \to 1$ and $\langle \sigma_x \rangle \to 0$. All of these phenomena are observed in Fig. \ref{Fig4}. The ground state energy $E_g$ along with the von Neumann entropy remains the same pre-rotation and post-rotation, which is illustrated in Fig. \ref{Fig5}a and c, respectively. \\

Intuitively, one can convert the post-rotation frame into the pre-rotation frame by replacing $\theta=\pi/3$ into the following rotation equations $\sigma_z = \tilde{\sigma}_z\cos\theta - \tilde{\sigma}_x\sin\theta$ and $\sigma_x = \tilde{\sigma}_z\sin\theta + \tilde{\sigma}_x\cos\theta$. In the pre-rotation frame, the physical picture is more sophisticated with a finite bias, tunnelling constant, diagonal and off-diagonal coupling. However, the known features from the post-rotation frame can be used to hypothesise the behaviour of $\langle \sigma_z \rangle$ and $\langle \sigma_x \rangle$ in the pre-rotation frame. As $\alpha \to 0$, $\langle \sigma_z \rangle \to -\sqrt{3}/2$ and $\langle \sigma_x \rangle \to 0.5$. After the transition point $\alpha_c=0.01475$, both $\langle \sigma_z \rangle$ and $\langle \sigma_x \rangle$ diverge into the degenerate ground states with one branch of $\langle \sigma_x \rangle \to \sqrt{3}/2$ and one branch of $\langle \sigma_z \rangle \to 0.5$. This rotation theory can be used to transform complicated physical pictures into simpler ones, allowing one to make accurate predictions without assumptions. Furthermore, the number of parameters is reduced, allowing for more efficient and faster computational time.\\

The competition between the diagonal and off-diagonal spin-bath coupling makes numerical studies of the ground-state properties in the anisotropic spin-boson model extremely challenging. A recent study \cite{noCRT} reported numerical variational calculations for the SBM under the rotating-wave approximation and the many challenges to achieve the true ground state. The novel rotational theory that bridges the pre- and post-rotation pictures of the SBM can be applied to obtain physical images more accurately for the model with simultaneous diagonal and off-diagonal coupling. \\

In the standard SBM, the localized-delocalized transition driven by the competition between the quantum tunneling and dissipation is well known to be of second order in the sub-Ohmic regime ($s<1$). With the non-vanishing spin bias, the quantum phase transition does not exist any more since the $Z_2$ symmetry is broken by the bias. With the rotational theory presented here, the model with any diagonal and off-diagonal coupling can be exactly mapped to the usual SBM. Hence, the generalized spin-boson model has the continuous transition only when the model parameters obey the relation in Eq. (27). Otherwise, the modified spin bias is non-vanishing, causing the spin to be trapped in the localized phase which in turn leads to no transition. It is in contrast to the conclusion in earlier numerical work \cite{Lu_Duan_Li_2013} obtained by the single Davydov D$_1$ \textit{Ansatz} where the emergence of a discontinuous first order phase transition was reported upon incorporation of the off-diagonal coupling. It further pushes the notion that only the variational wave function comprising of coherent superposition is valid in treating the ground-state transitions in the spin-boson model. 

\section{Conclusions}

Firstly, the multi-D$_2$ \textit{Ansatz} with time-independent variation has been shown to be an accurate, efficient numerical method in studying the diagonally coupled spin-boson model as the phase transition points yielded are in good agreement with those from other leading computational methods. This is evident in Table \ref{table1}, where results are displayed from the study conducted in the deep sub-ohmic regime with $s \in [0.1,0.5]$.\\

Secondly, the multi-D$_2$ \textit{Ansatz} is recruited to investigate other variants of the spin-boson models. In a two-bath SBM, for example, first order phase transitions are revealed regardless of ohmicity. In Ref.~\cite{Zhao_Zhao_2014}, when the ohmicity of the two baths are symmetric $(s=\bar{s})$ and non-symmetric $(s \neq \bar{s})$, DMRG appears to have falsely illustrated second order phase transitions in both cases with continuous behaviour in $\langle \sigma_x \rangle$, $\langle \sigma_z \rangle$ and $S_{V-N}$. A subsequent study employing the multi-D$_1$ \textit{Ansatz} \cite{Zhou_Chen_2014}, conducted shortly thereafter, unveiled first order transitions in both contexts as well which supports our findings in Sec.~\ref{ss_B}. It is also found that the DMRG results are susceptible to changes in the boson number, the Wilson chain length and the cutoff dimension of the matrices which could explain the inaccurate type of transition recorded. In this context, the multiple Davydov Ans\"{a}ze have demonstrated superior performance.\\

The rotational theory can be applied to SBM variants, where the individual modes of a bath are simultaneously diagonally ($\alpha$) and off-diagonally ($\beta$) coupled to the spin. In such systems with finite values of $\alpha$ and $\beta$, a rotational operator can be utilized to transform the Hamiltonian such that only a modified diagonal coupling remains ($\tilde{\beta} = 0$). This converts a sophisticated pre-rotation picture into a more widely studied post-rotation picture where the ground state properties, such as the quantum phase transition, are well known. In addition, results that have been already computed in the literature can be converted into its corresponding pre-rotation frame at no additional computational costs. For example, Zhang \textit{et al.} \cite{CSA_CQH} studied the diagonally coupled SBM with varying $\tilde{\Delta}$ and $s$ with the spin bias set to zero ($\tilde{\varepsilon}=0$). By selecting an appropriate $\theta$, the rotational theory can be used to transform the results into a Hamiltonian with finite $\alpha$, $\beta$, $\Delta$ and $\varepsilon$. 
As $\tilde{\Delta}$ increases, $|\epsilon|$ and $|\Delta|$ increases in the pre-rotation frame. A three-dimensional phase diagram of $\Delta$, $\varepsilon$ and $\alpha$ can then be illustrated. In another study \cite{K_Hur}, the case of a fixed finite tunnelling amplitude $\tilde{\Delta}$ with different detuning $\tilde{\varepsilon}$ is uncovered using the Bethe \textit{Ansatz}  techniques and NRG. It is shown that with increasing $\tilde{\varepsilon}$, the abrupt jump in $\langle \sigma_z \rangle$ at the critical transition point softens into a smooth crossover. This phenomenon can also be investigated in a SBM with both diagonal and off-diagonal coupling to the same mode, applying the rotational theory to the existing results. In summary, the rotational theory presented here is broadly applicable and independent of the specific numerical method used. Moreover, it enables more efficient and cost-effective computations by reducing the number of parameters required. Work is currently underway on extending the rotational theory to dynamical quantum phase transitions and interesting phenomena might be uncovered.

\section*{ACKNOWLEDGEMENTS}
One of us (Y.~Zhao) thanks Q.H.~Chen for useful discussion. Support from the Singapore Ministry of Education Academic Research Fund Tier 1 (Grant No.~RG2/24) and the Nanyang Technological
University ``URECA" Undergraduate Research Programme is gratefully acknowledged. 
N.~Zhou would also like to acknowledge support from National Natural Science Foundation of China under (Grant No.~11875120).

\section*{AUTHOR DECLARATIONS}
\subsection*{Conflict of Interest}
The authors have no conflicts to disclose.
\section*{DATA AVAILABILITY}
The data that support the findings of this study are available from the corresponding author upon reasonable request.

\end{document}